\pdfoutput=1
\documentclass[journal]{IEEEtran}
\usepackage{graphicx,amsmath,subfigure,epstopdf,color}
\usepackage{amsmath,graphicx,siunitx}

\usepackage{fancyhdr}
\renewcommand{\footskip}{20pt}
\begin{document}
%
\title{3-D Numerical Simulations of Twisted Stacked Tape Cables}
%
%
%

\author{Philipp A. C. Kr\"uger, Victor M. R. Zerme\~no, Makoto Takayasu, and Francesco~Grilli
\thanks{P. A. C.  Kr\"uger, V. Zerme\~no, and F. Grilli are with the Karlsruhe Institute of Technology, Karlsruhe Germany. M. Takayasu is with the Massachusetts Institute of Technology, Cambridge, MA, USA. Corresponding author's e-mail: francesco.grilli@kit.edu.}
\thanks{This work was partly supported by the Helmholtz Association (Grant VH-NG- 617).}
\thanks{Manuscript received August 10, 2014.}}

\maketitle

\begin{abstract}
Different magnet applications require compact high current cables. Among the proposed solutions, the Twisted Stacked Tape Cable (TSTC) is easy to manufacture and has very high tape length usage efficiency. In this kind of cables the tapes are closely packed, so that their electromagnetic interaction is very strong and determines the overall performance of the cable. Numerical models are necessary tools to precisely evaluate this interaction and to predict the cable's behavior, e.g. in terms of effective critical current and magnetization currents. For this purpose, we  developed a fully three-dimensional model of a TSTC, which not only takes into account the twisted geometry of these cables, but is also able to account for the contact resistances of the current terminations. The latter can have profound influence on the way the current is partitioned among the tapes, especially on short laboratory prototypes.

In this paper, we first use the numerical model to compute the critical current and the magnetization AC loss of a twisted tape, showing the differences with the case of a straight tape. Then, we use it to calculate the current distribution in a TSTC cable, comparing the results with those experimentally obtained on a cable composed of four straight stacked tapes.
The results show the ability of the model to simulate twisted conductors and constitutes a first step toward the simulation of TSTC in high-field magnet applications. The presented modeling approach is not restricted to the TSTC geometry, but may be used for any cable configuration with periodical translational symmetry.

\end{abstract}
\section{Introduction}
\IEEEPARstart{D}{ifferent} 
cable designs are currently being considered for fusion magnet applications using HTS coated conductors~\cite{Fietz:FED13}. In all these designs, the tapes are very compactly arranged, in a geometrical configuration involving either twisting, transposition or helical winding. Their electromagnetic interaction is very strong, and it is therefore desirable to have numerical tools able to predict the performance of such cables, for example the calculation of the effective critical current, the current distribution between the tapes, and their losses when subjected to varying currents and fields. Such tools must be able to calculate the local components of the magnetic field: in this way, the typically anisotropic transport properties of HTS tapes can be properly taken into account, as it has been done by a variety of 2-D models developed by several authors in the past years (see, for example,~\cite{Amemiya:PhysC01, Stavrev:TAS02, Pardo:SST11, Gomory:TAS13, Grilli:TAS13}).

In this paper, we developed a full 3-D model for twisted superconductors based on the popular $H$-formulation of eddy currents problems implemented in the software package Comsol~\cite{Hong:SST06, Brambilla:SST07}, which has been recently extended to 3-D~\cite{Zhang:SST12, Grilli:Cryo13, Zermeno:SST14}. In addition to considering twisted structures and including the field dependence of $J_c$, the model is able to simulate the contact resistances of the electrical terminations, while keeping the possibility of simulating only a periodic cell (e.g. one twist pitch) of the superconducting parts.

We use this model to simulate the electromagnetic behavior of a twisted tape and of a Twisted-Stacked-Tape-Cable in a variety of working conditions, emphasizing the effects of the twisted geometry on the tape's and cable's performances. We also use it to study the current distribution between tapes in a straight stacked-tape cable manufactured at MIT~\cite{Takayasu:SST12}, for which experimental data are available.  

In this work, we use critical current density values, $J_c$, typical for coated conductors at \SI{77}{\kelvin}, also for the purpose of validations against available experimental data. The applied background fields are rather small, in the range of 5-\SI{20}{\milli\tesla}. The main goal of this paper is to show the feasibility of full 3-D electromagnetic calculations involving twisted coated conductors and assemblies thereof. The paper can therefore be seen as a first stepping stone toward the simulation of  specific applications of these conductors (e.g. magnets for fusion or accelerator applications) characterized by much higher currents and background fields as well as by different angular dependences of $J_c$.

\section{Single Twisted Tape}
First we simulated a single twisted coated conductor tape, \SI{4}{\milli\meter} wide, with a twist pitch of 40 cm. The superconductor is modeled with a power-law resistivity which includes the dependence of the critical current density on the components of the magnetic flux density parallel and perpendicular to the flat face of the tape~\cite{Kario:SST13, Grilli:TAS14c}:
\begin{equation}\label{eq:JcB}
J_c(B_\parallel,B_\perp)=\frac{J_{c0}}{\left [1+\sqrt{(k B_\parallel)^2+B_\perp^2}\,/B_c\right ]^b}.
\end{equation}
Here we  considered values typical of the characterization of HTS  coated conductors at \SI{77}{\kelvin}: $J_{c0}$=\SI{21.2}{\giga\ampere\per\square\meter}, $B_c$=\SI{32.5}{\milli\tesla}, $k$=0.275, $b$=0.6, power index $n$=21.

In the finite-element (FE) model, the geometry is built by first drawing the transversal cross-section of the tape ($xy$ plane) and then by extruding it in several steps along the tape direction ($z$ direction). This allows fixing the mesh of the cross-section  and sweeping it along the direction of extrusion~\cite{Krueger:Thesis14}. 
The local parallel and perpendicular field components appearing in~\eqref{eq:JcB} need to be locally evaluated at every point of the tape geometry.
For the  general case of simultaneous presence of transport current and external field, it is necessary to simulate one full twist pitch. Other situations (e.g. external field only) may allow reducing the simulated length to half or a quarter of the pitch length, 

\subsection{$E-I$ Characteristics with Background Field}

For our first tests, we calculated the voltage-current characteristics of a twisted tape in a uniform magnetic field of \SI{20}{\milli\tesla} applied along the $y$-direction. For this purpose, the external field is increased from zero to the maximum value with a linear ramp of $t_H$=\SI{1}{\second}; then it is left constant, and, starting at $t_I$=\SI{10}{\second}, the current is ramped up to \SI{100}{\ampere} with a rate of \SI{10}{\ampere\per\second}).
\begin{figure}[t!]
\begin{center}
\includegraphics[width=\columnwidth]{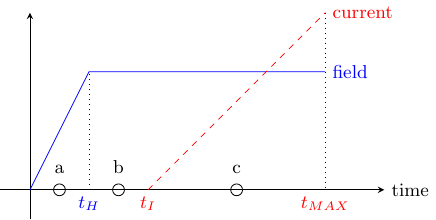}
\end{center}
\caption{\label{fig:ramps_plot}Magnetic field and current ramps applied for the determination of the tape's or cable's critical current in the presence of a background field. The circles represent the instants when the current density distributions of Fig.~\ref{fig:123} are taken.}
\end{figure}
\begin{figure}[t!]
\begin{center}
\includegraphics[width=\columnwidth]{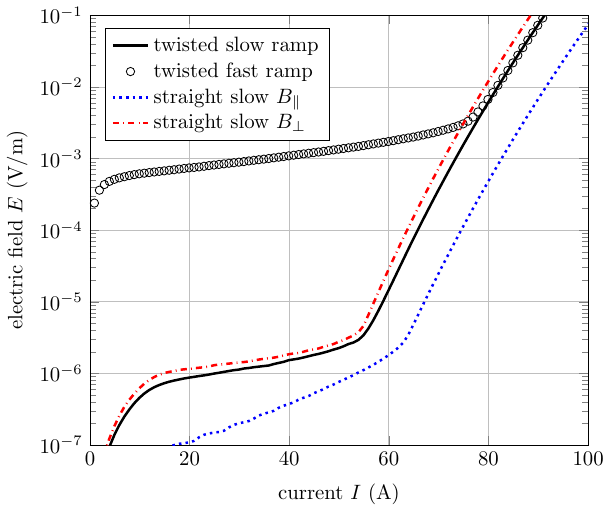}
\end{center}
\caption{\label{fig:EI_single_plot}$E-I$ characteristic of a twisted tape with a background field, compared to those of an equivalent straight tape experiencing a purely perpendicular or parallel background field. The curve obtained with ``fast'' field and current ramps is also shown (circle symbols).}
\end{figure}
\begin{figure}[t!]
\begin{center}
\includegraphics[width=\columnwidth]{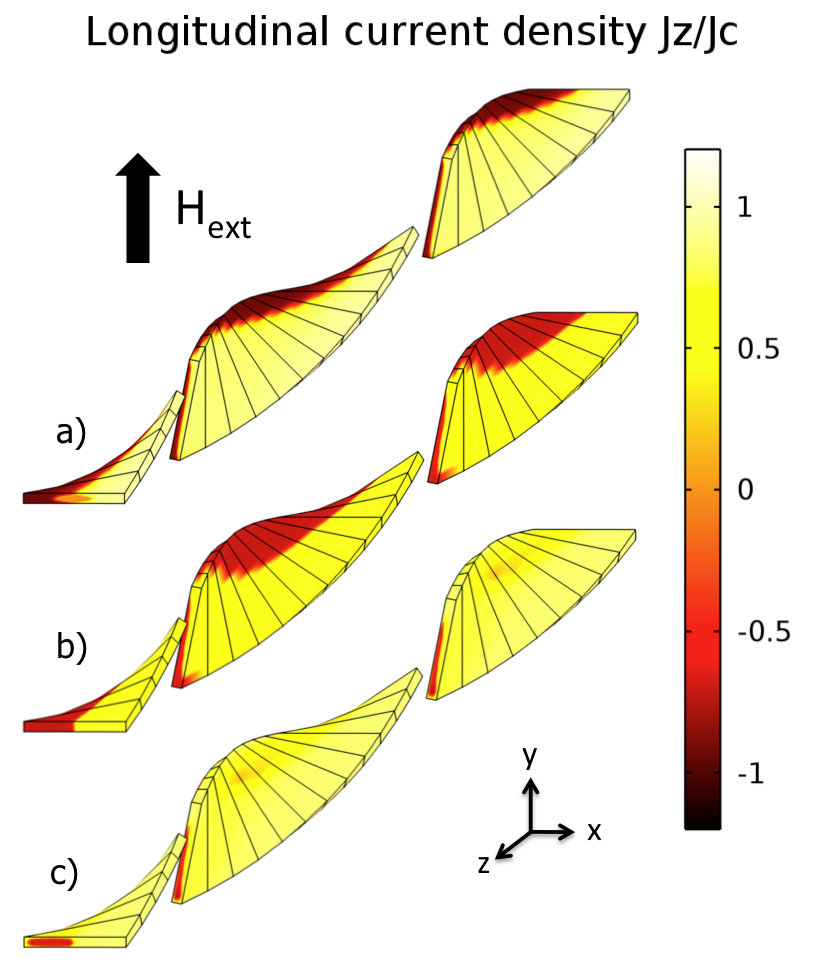}
\end{center}
\caption{\label{fig:123}Distribution of the current density component $J_z$ (normalized to the local $J_c$) at different time instants of the field and current ramps (see circles in Fig.~\ref{fig:ramps_plot}). Some portions of the twisted tape are removed, in order to make the internal current distribution  at the different places visible. For readability purposes, a thick (\SI{400}{\micro\meter}) tape is shown here, but similar distributions were found for a tape as thin as \SI{10}{\micro\meter}.}
\end{figure}
The applied field and current ramps are schematically shown in Fig.~\ref{fig:ramps_plot} and the resulting $E-I$ characteristic is given by the continuous line in Fig.~\ref{fig:EI_single_plot} (``twisted slow ramp'' in the legend), where $E$ is the volumetric average of $E_z$. For the twisted tape, the critical current (determined when the electric field reaches \SI{1e-4}{\volt\per\meter}) is \SI{66}{\ampere}. In the case of a straight tape (with the same $J_c(B)$ properties) subjected to a purely perpendicular or parallel field, the critical current is 63.5 and \SI{73.5}{\ampere}, respectively, as a consequence of a stronger or weaker reduction of $J_c(B)$ according to~\eqref{eq:JcB}. The twisted tape experiences both parallel and perpendicular local components, so its critical current is between the two cases of the straight tape in perpendicular and parallel applied field. 

Besides the effects related to the presence of a twisted geometry discussed just above, a closer look at the current density distribution inside the tape at different instants during the field and current ramps reveals a general important consequence of the use of a power-law (PL) in transient electromagnetic models -- not particularly related to twisted geometries: current relaxation. During the initial field ramp ($t<t_H$), the current density has a pattern very similar to that described by the critical state model (CSM): at the edges of the tape there are regions carrying the maximum possible current (in the framework of the CSM, $J/J_c=1$, here the current density is slightly over-critical because of the PL) -- see Fig.~\ref{fig:123}(a). As the field is held constant ($t_H < t < t_I$), the current density relaxes to sub-critical values (Fig.~\ref{fig:123}(b)). Finally, as the new current kicks in ($t>t_I$), the regions with higher $J_c$ values change: note that the current distribution is no longer symmetric because of the superposition of the transport current to the pre-existing magnetization currents (Fig.~\ref{fig:123}(c)).
Whether this relaxation behavior is realistic or just a consequence of the use of the PL relationship is not clear. In~\cite{Lahtinen:JAP14}, the authors compare models based on the PL and on the CSM to experiments for the calculation of AC ripple losses and argue that the behavior observed in experiments is in better agreement with the predictions based on the CSM, as opposed to the PL model. However, they also say that further investigations are necessary on the topic.
With the PL, a critical state-like behavior can be obtained by quickly ramping current and fields (e.g. with $t_H$=$t_I$=\SI{0.01}{\second} and $t_{MAX}$=\SI{0.02}{\second} -- see Fig.~\ref{fig:ramps_plot} for reference). In Fig.~\ref{fig:ramps_plot}, the circle symbols (``twisted fast ramp'' in the legend) represent the $E-I$ characteristic obtained with such fast ramp. In that case, the obtained $E-I$ curve is different from the one that can be obtained in real experiments (compare also with section~\ref{sec:contact} and with the results reported in the appendix of~\cite{Grilli:TAS14c}), but the correct value of the critical current can be obtained by extrapolating the upper-right part of the curve down to the critical electric field of \SI{1e-4}{\volt\per\meter}. The advantage of doing so is that  `fast' ramps are much faster to simulate in transient electromagnetic problems. 
The reason is that when using the $H$-formulation and the PL relationship, the zero divergence of the magnetic flux density is only imposed by means of the time derivative term in the resulting non-linear diffusion equation. This is explained in detail in~\cite{Zermeno:JAP13} and in section 2 of~\cite{Zermeno:SST13}. For low frequencies, this time derivative term becomes numerically less significant. Overall, this increases the workload on the solver which requires more iterations for reaching a solution in accordance with the target tolerance values.

\subsection{Magnetization AC Losses}
\begin{figure}[t!]
\begin{center}
\includegraphics[width=\columnwidth]{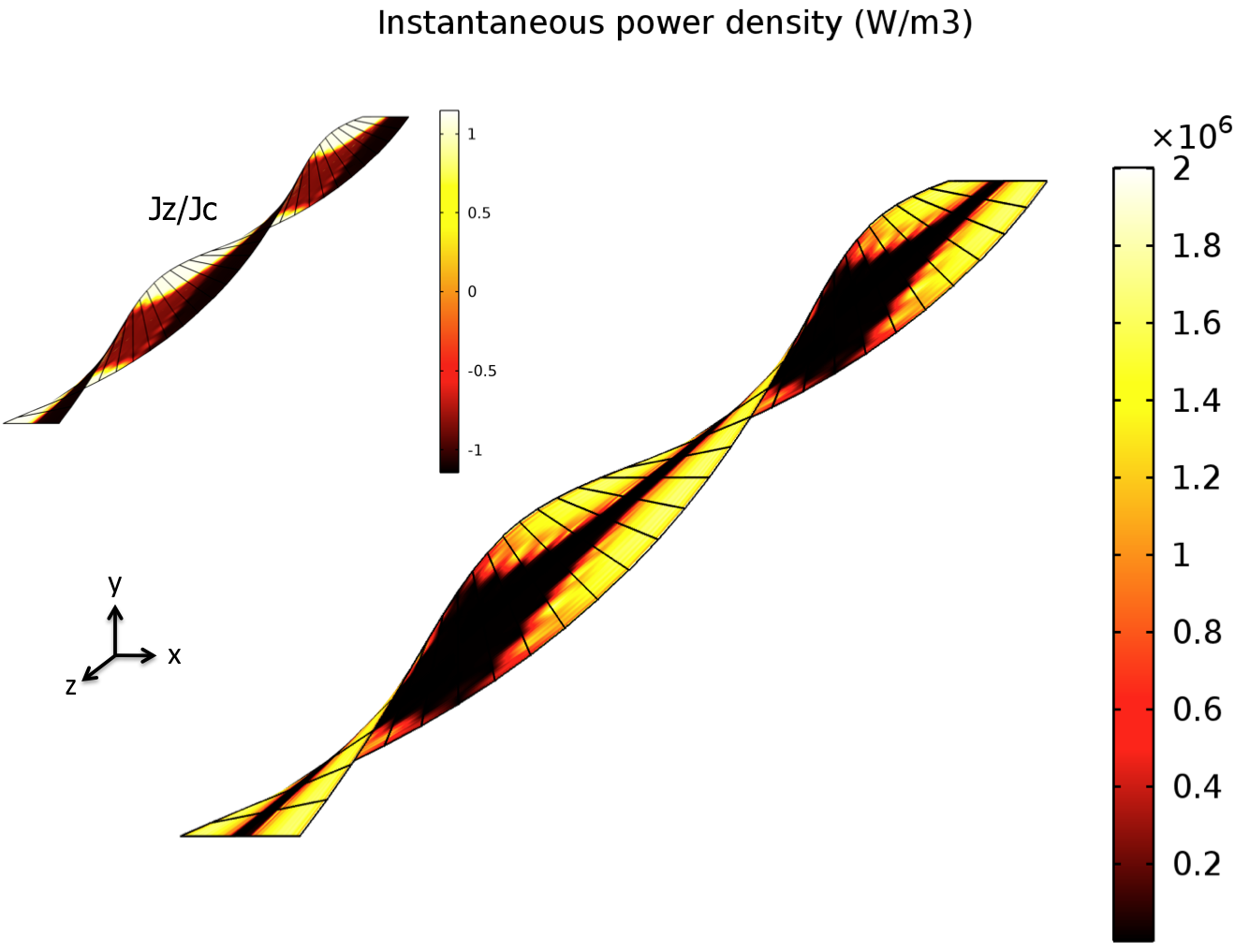}
\end{center}
\caption{\label{fig:JP}Instantaneous power density in a twisted tape subjected to an oscillating (frequency \SI{50}{\hertz}) magnetic field of \SI{20}{\milli\tesla}. The corresponding current density distribution is shown in the inset. The figures refer to the instant of the peak of the field.}
\end{figure}

As a second step, we calculated the magnetization AC losses caused by a background magnetic field oriented along the $y$ direction. The field has a varying orientation with respect to the twisted tape, whose behavior varies from `strip-like' (when the field is perpendicular to the flat face of the tape) to `slab-like' (when the field is parallel). This is clearly visible in Fig.~\ref{fig:JP}: shown are the current density distribution (along $z$) induced by a  field of \SI{20}{\milli\tesla} of amplitude applied in the $y$-direction and the corresponding power density, which is much higher in the regions of the tape characterized by a strip-like behavior.

\begin{figure}[t!]
\begin{center}
\includegraphics[width=\columnwidth]{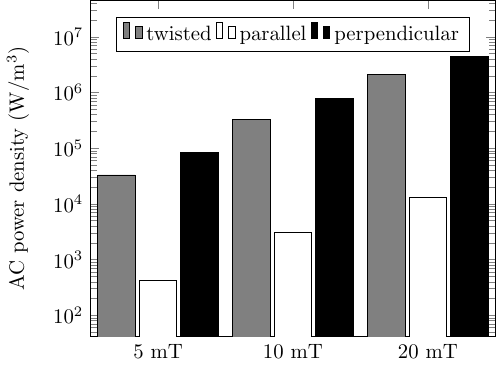}
\end{center}
\caption{\label{fig:AC_losses_bar}
AC power density for the case of a tape subjected to a background field (at \SI{50}{\hertz})  of different amplitudes. Comparison between a twisted tape and a straight tape experiencing only a perpendicular or parallel field.}
\end{figure}

The power density for different field amplitudes is displayed in Fig.~\ref{fig:AC_losses_bar}: as for the critical currents (Fig.~\ref{fig:EI_single_plot}), the power density of the twisted tape lie between that of a similar straight tape experiencing fully perpendicular and fully parallel magnetic field. The main difference with the critical current case discussed above is that the AC loss dynamics are very different in strip-like~\cite{Brandt:PRB93, Zeldov:PRB94} and slab-like~\cite{Wilson83} configurations, which results in power densities orders of magnitude apart, as shown in the figure. The losses of a straight tape in perpendicular and parallel field can be calculated with simple 2-D models. However, they  constitute just two far limit values and cannot be used for a realistic evaluation of the dissipation in a twisted tape. This is a clear example of the usefulness of a numerical model able to calculate the losses of twisted tapes.

\section{Twisted Stacked-Tape Cable}
\begin{figure}[t!]
\begin{center}
\includegraphics[width=\columnwidth]{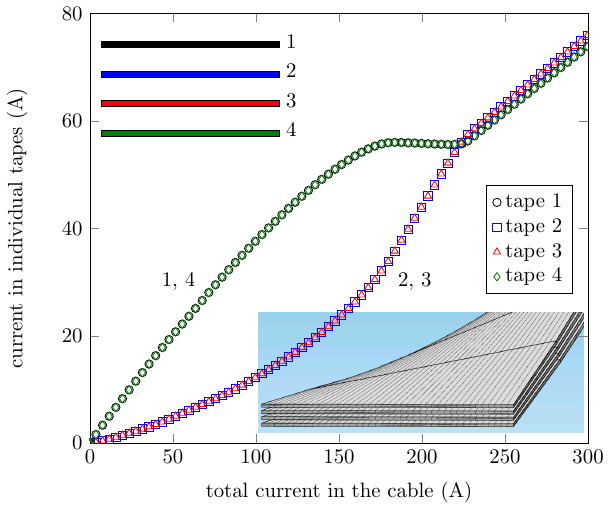}
\end{center}
\caption{\label{fig:4tapes}Distribution of the current in the 4-tape TSTC with an exemplary image of the utilized mesh.}
\end{figure}

We considered a twisted stacked-tape cable composed of four tapes, subjected to a current ramp in background field. In order to construct the geometry and  the mesh, we  followed the same procedure as in the case of the single twisted tape: the transversal cross section is first built in a 2-D work plane and successively extruded along the cable direction with a multi-step process. The current is imposed to the whole superconductor cross-section and let free to distribute in the different tapes. Additionally, we imposed that the current entering each tape's cross section is the same at the two ends of the simulated geometry (one twist pitch). The results of the current distribution between the tapes are shown in Fig.~\ref{fig:4tapes}. At the beginning, most of the current flows in the most external tapes. When they reach the current capacity, significant current starts to flow in the internal tapes as well. For very high total current, all the tapes carry almost the same current: however, the innermost tapes carry a little more current than the outermost ones, most probably because they have a slightly higher critical current as a consequence of the smaller experienced magnetic field. Figure~\ref{fig:4tapes} also shows the mesh for a portion of the TSTC.

\section{Influence of Contact Resistances}\label{sec:contact}

\begin{figure}[t!]
\begin{center}
\includegraphics[width=\columnwidth]{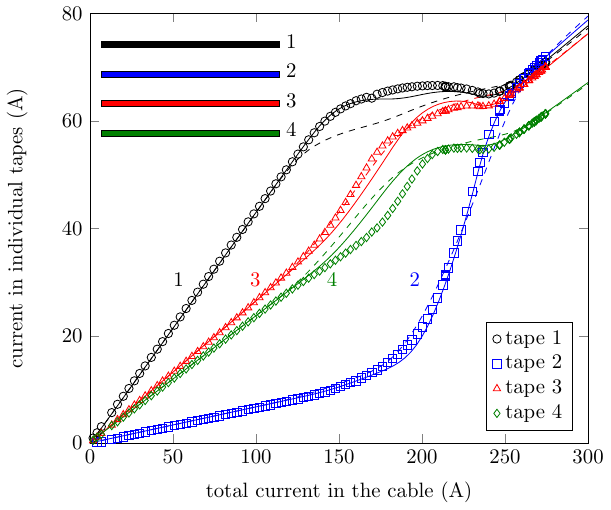}
\end{center}
\caption{\label{fig:ASC_straight_plot}
Measured (symbols) and calculated (lines) distribution of the current in a stacked-tape cable composed of four tapes. The continuous and dashed lines represent the results of simulation with and without angular dependence of $J_c$, respectively.}
\end{figure}
In addition to being able to simulate twisted tapes, our model is able to take into account the contact resistance offered by the electrical terminations through which the current is injected. This is done by placing the resistances in a simulation domain disconnected from that of the tapes and by transferring the current from a contact resistance to the corresponding tape by means of integral constraints. The advantage of this approach with respect to simulating the full geometry is that, for the simulation of the superconducting tapes, one can take full advantage of the existing periodicity. So, for example, one can still simulate just a twist pitch of superconducting cable and adjust the contact resistance to take into account the real cable length. 
Technical details for the implementation of this approach can be found in~\cite{Krueger:Thesis14}.

Here we used the model to calculate the current distribution in  a straight cable composed of four stacked tapes, for which experimental data are available~\cite{Takayasu:SST12}.
The results are shown in Fig.~\ref{fig:ASC_straight_plot}. The current distribution is different from that presented in Fig.~\ref{fig:4tapes} in two points: 1) The initial distribution of the current is governed by the contact resistance of each tape; 2) In the high current regime, the currents carried by the tapes are different: in particular, tape number 4 carries substantially less current than the others because it has a much lower critical current than the others.

One similarity with the data of Fig.~\ref{fig:4tapes} is that in some tapes the current does not increase monotonically, but it shows a plateau or even a decrease. This is probably an effect of the generated self-field: as the current carried by the cable increases, so does the generated self-field, which in turns decreases the maximum current a given tape can carry. It is worth noting that this effect is not present if the $J_c(B)$ dependence of the superconductor is not taken into account (dashed lines in the figure). 

More details can be found in a recent publication of ours~\cite{Zermeno:Bratislava14}, where other techniques to take the presence of contact resistances into account are presented. The current approach, however, is the only one that allows simulating contact resistances and complex 3-D geometries (such as TSTC) within the same model.

\section{Conclusion}
In this paper we presented preliminary simulation results of individual and stacked twisted HTS tapes, both in DC and AC conditions.
For the evaluation of the critical current of the tapes or cables in background magnetic field, the issue on the type of field/current ramps to apply with a PL model is discussed. Fast ramps, although not describing the behavior occurring in reality, can provide a fast calculation of the effective $I_c$.
The 3-D model offers the possibility of including the presence of contact resistances, which can significantly influence the current distribution between the tapes. The simulation domain of the contact resistance is disconnected from that of the superconductor devices, which allows the simulation of translational geometries without the introduction of end-effect perturbations.
The model presented here has all the necessary features to handle simulation of HTS devices employing twisted conductors with complex angular $J_c(B)$ dependence. In future work it will be used to investigate the electromagnetic behavior of TSTC conductors experiencing the conditions typical of large-scale applications, such as fusion or accelerator magnets.


\end{document}